\documentstyle [12pt]{article}
\makeatletter \oddsidemargin  0in \evensidemargin 0in
\textwidth16cm \RequirePackage[dvips]{graphicx} \textheight 20.5cm
\newcommand{\singlespacing}{\let\CS=\@currsize\renewcommand{\baselinestretch}{1.5}\tiny\CS}
\makeatletter \oddsidemargin  -.1in \evensidemargin -.1in
\newcommand{\doublespacing}{\let\CS=\@currsize\renewcommand{\baselinestretch}{1.35}\tiny\CS}

\def\@citex[#1]#2{\if@filesw\immediate\write\@auxout{\string\citation{#2}}\fi
  \def\@citea{}\@cite{\@for\@citeb:=#2\do
    {\@citea\def\@citea{,\linebreak[0]\hskip0pt plus .2em}%
      \@ifundefined{b@\@citeb}%
    {{\bf ?}\@warning{Citation `\@citeb' on page \thepage\space undefined}}%
      \hbox{\csname b@\@citeb\endcsname}}}{#1}}

\newtheorem{rule-def}[theorem]{Rule}
\begin{document}
\title{\bf Secretly Broadcasting Five Qubit Entangled state among three parties from W- type states}\author{I.Chakrabarty $^{1,2}$\thanks{Corresponding author:
E-Mail-indranilc@indiainfo.com }, B.S.Choudhury $^2$ \\
$^1$ Heritage Institute of Technology,Kolkata-107,West Bengal,India\\
$^2$ Bengal Engineering and Science University, Howrah, West
Bengal, India }
\date{}
\maketitle{}
\begin{abstract}
In this work we investigate the problem of secretly broadcasting
five qubit entangled state between three different partners We
implement the protocol described in ref [16] on three particle
W-state shared by three distant partners Alice,Bob and Charlie.
The problem is interesting in the sense it is the first attempt
to broadcast five qubit entangled state between three parties.
\end{abstract}
\section{Introduction}
Linearity of quantum theory doesn't allow us to amplify and
delete an arbitrary quantum state [1][6]. Although nature prevents
us from amplifying an unknown quantum state but nevertheless one
can always design a quantum cloning machine that duplicates an
unknown quantum state
with a fidelity less than unity [1,2,3,4,5].\\
However, the authors in [7] if the states are linearly
independent, they do can be cloned by a unitary-reduction
process. In the past years,much progress has been made in
designing quantum cloning machine. Buzek-Hillery took the first
step towards the construction of approximate quantum cloning
machine [2]. This machine is known as universal quantum cloning
machine (UQCM)as the quality of the copies produced by their
machine remain same for all input state. Later Gissin -Massar
showed the machine to be optimal [3]. After that the different
sets of quantum cloning machines like the set of universal
quantum cloning machines, the set of state dependent quantum
cloning machines (i.e. the quality of the copies depend on the
input state) and the probabilistic quantum cloning machines were
proposed. Entanglement[8] plays a crucial role in computational
and communicational purposes and is used as a valuable resource in
quantum information processing, quantum entanglement quantum
cryptography [9,10],quantum super dense coding [11] and quantum
teleportation [12]. An interesting feature of quantum information
processing is that information can be "encoded" in non-local
correlations between two separated particles.  "Pure" quantum
entanglement, is more "valuable". Therefore, it becomes
interesting  to extract pure quantum entanglement from a
partially entangled state [10]. In other words, it is possible to
compress locally an amount of quantum information. Now one can
ask an question : whether the opposite is true or not i.e. can
quantum correlations be "decompressed"? This question was tackled
by several researchers using the concept of "Broadcasting of
quantum inseparability". Broadcasting is nothing but a local
copying of non-local quantum correlations. That is the
entanglement originally shared by a single pair is transferred
into two less
entangled pairs using only local operations.\\\\
Suppose two distant parties A and B share two qubit-entangled
state
\begin{eqnarray}
|\psi\rangle_{AB}=\alpha|00\rangle_{AB}+\beta|11\rangle_{AB}
\end{eqnarray}

The first qubit belongs to A and the second belongs to B. Each of
the two parties now perform local copier on their own qubit and
then the input entangled state ø has been broadcast if for some
values of the probability $\alpha^2$\\
 (1) non-local output states are
inseparable, and \\
 (2) local output states are separable.\\
V.Buzek et.al. showed that the decompression of initial quantum
entanglement is theoretically possible, i.e. if we start with an
entangled pair, then two less entangled pairs can be obtained by
local operations. That means inseparability of quantum states can
be partially broadcasted (cloned) with the help of local
operation. They used optimal universal quantum cloners for local
copying of the subsystems and showed that the non-local outputs
are inseparable if $\alpha^2$ lies in the interval
$(\frac{1}{2}-\frac{\sqrt{39}}{16},\frac{1}{2}+\frac{\sqrt{39}}{16})$.\\
Further S.Bandyopadhyay et.al. [13]  showed that only those
universal quantum cloners whose fidelity is greater than
$\frac{1}{2}(1 + \sqrt{\frac{1}{3}} )$ are suitable because only
then the non-local output states becomes inseparable for some
values of the input parameter $\alpha$ and also proved that an
entanglement is optimally broadcast only when optimal quantum
cloners are used for the purpose of local copying . They also
showed that broadcasting of entanglement into more than two
entangled pairs is not possible using only local operations.
I.Ghiu investigated the broadcasting of entanglement by using
local $1\rightarrow 2$ optimal universal asymmetric Pauli
machines and showed that the inseparability is optimally
broadcast when symmetric cloners are applied [14]. Few years back
we study the problem of broadcasting of entanglement using state
dependent quantum cloning machine as a local copier. We show that
the length of the interval for probability-amplitude-squared for
broadcasting of entanglement using state dependent cloner can be
made larger than the length of the interval for
probability-amplitude-squared for broadcasting entanglement using
state independent cloner. In that work we showed that there
exists local state dependent cloner which gives better quality
copy (in terms of average fidelity)
of an entangled pair than the local universal cloner [15].\\
In recent past Adhikari et.al in their paper [16] showed that
secretly broadcasting of three-qubit entangled state between two
distant partners with universal quantum cloning machine is
possible. They generalized the result to generate secret
entanglement among three parties. Recently Adhikari et.al in ref
[17] proposed a scheme for broadcasting of continuous variable
entanglement. In a recent work we have presented a scheme of broadcasting
W state secretly between three distant partners [18]. \\
Motivated by ref [16]we investigate the problem of secretly
broadcasting Five qubit entangled state between three distant
Alice, Bob and Charlie. In this work we consider a W type of state
\begin{eqnarray}
|X\rangle_{123}=\alpha|001\rangle_{123}+\beta|010\rangle_{123}+\gamma|100\rangle_{123}
\end{eqnarray}
The three parties then apply optimal universal quantum cloning
machine on their respective qubits to produce six qubit state.One
of the party (say, Alice) then performs measurement on her quantum
cloning machine state vectors.  Bob also performs measurement on
his quantum cloning machine state vectors . Charlie also does
measurement on his cloning state vectors. Each party inform
others about their measurement result using Goldenberg and
Vaidman's quantum cryptographic scheme [20] based on orthogonal
state. Since the measurement results are interchanged secretly so
Alice ,Bob and Charlie share secretly six qubit state. They again
apply the cloning machine on one of their respective qubits and
generate nine qubit state. Now once again each of them do
measurement on their machine state vectors and  secretly inform
each other about measurement outcomes. Interestingly, we find a
five qubit entangled state between the original qubit of Alice
and two cloned copies from each of Bob's and Charlie's subsystem.
We observe that the local output states are also separable. Thus
we able to broadcast a five qubit entangled state among three
distant partners. Since three parties perform measurements twice
on their machine state vectors and communicate with each other
secretly, so the final five qubit entangled state shared by them
can be used as a secret quantum channel between these parties.
Any fourth party have to obtain the information about the
measurement results in order to know about the five qubit
entangled state.\\

In broadcasting of inseparability, we generally use
Peres-Horodecki criteria [21,22] to show the inseparability of
non-local
outputs and separability of local outputs.\\\\
\textbf{Peres-Horodecki Theorem :}The necessary and sufficient
condition for the state $\rho$ of two spins $\frac{1}{2}$ to be
inseparable is that at least one of the eigen values of the
partially transposed operator defined as
$\rho^{T}_{m\mu,n\nu}=\rho_{m\mu,n\nu}$, is negative. This is
equivalent to the condition that at least one of the two
determinants\\\\
$W_{3}= \begin{array}{|ccc|}
  \rho_{00,00} & \rho_{01,00} & \rho_{00,10} \\
  \rho_{00,01} & \rho_{01,01} & \rho_{00,11} \\
  \rho_{10,00} & \rho_{11,00} & \rho_{10,10}
\end{array}$ and $W_{4}=\begin{array}{|cccc|}
   \rho_{00,00} & \rho_{01,00} & \rho_{00,10} & \rho_{01,10}\\
  \rho_{00,01} & \rho_{01,01} & \rho_{00,11} & \rho_{01,11} \\
  \rho_{10,00} & \rho_{11,00} & \rho_{10,10} & \rho_{11,10} \\
  \rho_{10,01} & \rho_{11,01} & \rho_{10,11} & \rho_{11,11}
\end{array}$\\
is negative.\\
The protocol is interesting in the sense that, this is a first
attempt to broadcast five qubit entangled state from a three
qubit W-state using  B-H local quantum copying machine.
\section{Local Copying of W-type state and Broadcasting}
Let three parties Alice, Bob  and Charlie share a W-type state of
the form
\begin{eqnarray}
|X\rangle_{123}=\alpha|001\rangle_{123}+\beta|010\rangle_{123}+\gamma|100\rangle_{123}
\end{eqnarray}
where without any loss of generality, we have assumed that
$\alpha,\beta,\gamma$ are all real with
$\alpha^2+\beta^2+\gamma^2=1$. The qubits 1,2,3 are in possession
with Alice,Bob and Charlie respectively. \\
The B-H cloning transformation is given by,
\begin{eqnarray}
|0\rangle\longrightarrow\sqrt{\frac{2}{3}}|00\rangle|\uparrow\rangle+\frac{1}{\sqrt{6}}(|01\rangle+|10\rangle)|\downarrow\rangle\nonumber\\
|1\rangle\longrightarrow\sqrt{\frac{2}{3}}|11\rangle|\downarrow\rangle+\frac{1}{\sqrt{6}}(|01\rangle+|10\rangle)|\uparrow\rangle
\end{eqnarray}
where $\{|\uparrow\rangle,|\downarrow\rangle\}$ are orthogonal
quantum cloning machine state vectors.\\
Now Alice, Bob and Charlie apply the cloning machine defined by
equation (4) on their respective qubits and hence obtain six qubit
state given by,
\begin{eqnarray}
&&|X_1\rangle_{142536}={}\nonumber\\&&
\alpha[\sqrt{\frac{2}{3}}|00\rangle_{14}|\uparrow\rangle^A+\frac{1}{\sqrt{6}}(|01\rangle_{14}+|10\rangle_{14})|\downarrow\rangle^A]\otimes{}\nonumber\\&&
[\sqrt{\frac{2}{3}}|00\rangle_{25}|\uparrow\rangle^B+\frac{1}{\sqrt{6}}(|01\rangle_{25}+|10\rangle_{25})|\downarrow\rangle^B]\otimes{}\nonumber\\&&
[\sqrt{\frac{2}{3}}|11\rangle_{36}|\downarrow\rangle^C+\frac{1}{\sqrt{6}}(|01\rangle_{36}+|10\rangle_{36})|\uparrow\rangle^C]{}\nonumber\\&&
+\beta[\sqrt{\frac{2}{3}}|00\rangle_{14}|\uparrow\rangle^A+\frac{1}{\sqrt{6}}(|01\rangle_{14}+|10\rangle_{14})|\downarrow\rangle^A]\otimes{}\nonumber\\&&
[\sqrt{\frac{2}{3}}|11\rangle_{25}|\downarrow\rangle^B+\frac{1}{\sqrt{6}}(|01\rangle_{25}+|10\rangle_{25})|\uparrow\rangle^B]\otimes{}\nonumber\\&&[\sqrt{\frac{2}{3}}|00\rangle_{36}|\uparrow\rangle^C+\frac{1}{\sqrt{6}}(|01\rangle_{36}+|10\rangle_{36})|\downarrow\rangle^C]{}\nonumber\\&&+\gamma[\sqrt{\frac{2}{3}}|11\rangle{14}|\downarrow\rangle^A+\frac{1}{\sqrt{6}}(|01\rangle_{14}+|10\rangle_{14})|\uparrow\rangle^A]\otimes{}\nonumber\\&&
[\sqrt{\frac{2}{3}}|00\rangle_{25}|\uparrow\rangle^B+\frac{1}{\sqrt{6}}(|01\rangle_{25}+|10\rangle_{25})|\downarrow\rangle^B]\otimes{}\nonumber\\&&[\sqrt{\frac{2}{3}}|00\rangle_{36}|\uparrow\rangle^C+\frac{1}{\sqrt{6}}(|01\rangle_{36}+|10\rangle_{36})|\downarrow\rangle^C]
\end{eqnarray}
where qubits with subscripts'4','5','6' are cloned copies of the
qubits with subscripts'1','2','3'respectively.\\
Now if Alice, Bob, Charlie carry out measurements on their
machine state vectors
$\{|\uparrow\rangle^A,|\downarrow\rangle^A\}$,$\{|\uparrow\rangle^B,|\downarrow\rangle^B\}$,$\{|\uparrow\rangle^C,|\downarrow\rangle^C\}$
and exchange the measurement results with each other secretly
with the help of Goldenberg and Vaidman's quantum cryptographic
scheme based on orthogonal state [20].\\
The tensor products of machine state vectors after the
measurement is given by the following table.\\\\\\

{\bf TABLE 1:}\\
\begin{tabular}{|c|c|}
\hline Serial Number & Measurement Results   \\
\hline 1 &
$|\uparrow\rangle^A|\uparrow\rangle^B|\uparrow\rangle^C$\\
\hline 2 & $|\uparrow\rangle^A|\uparrow\rangle^B|\downarrow\rangle^C$ \\
\hline 3 & $|\uparrow\rangle^A|\downarrow\rangle^B|\downarrow\rangle^C$ \\
\hline 4
&$|\uparrow\rangle^A|\downarrow\rangle^B|\uparrow\rangle^C$
\\ \hline 5 & $|\downarrow\rangle^A|\uparrow\rangle^B|\uparrow\rangle^C$ \\
\hline 6 & $|\downarrow\rangle^A|\uparrow\rangle^B|\downarrow\rangle^C$\\
\hline 7 & $|\downarrow\rangle^A|\downarrow\rangle^B|\uparrow\rangle^C$\\
\hline 8 & $|\downarrow\rangle^A|\downarrow\rangle^B|\downarrow\rangle^C$\\
\hline
\end{tabular}\\
Now if we assume that the measurement outcome is
$|\uparrow\rangle^A|\uparrow\rangle^B|\uparrow\rangle^C$, then the
six qubit post measurement state shared by Alice, Bob and Charlie
is given by,
\begin{eqnarray}
 |X_2\rangle_{142536}=\frac{1}{\sqrt{N} }\{\alpha(|000001\rangle+|000010\rangle)+\beta(|000100\rangle+|001000\rangle)+\gamma(010000|\rangle+|100000\rangle)\}
\end{eqnarray}
where $N=\sqrt{2\alpha^2+2\beta^2+2\gamma^2}$ is the normalization constant.\\
 Now once again three
distant partners apply local cloning operations given by equaton
(4) on their qubits '4','5','6' respectively and performs
measurement in the machine state vectors. They also
exchange the measurement results once again (table1) using the same cryptographic scheme.\\
Again if we assume that the measurement outcome is
$|\uparrow\rangle^A|\uparrow\rangle^B|\uparrow\rangle^C$, then the
nine qubit state shared by Alice, Bob and Charlie is given by,
\begin{eqnarray}
|X_3\rangle_{147258369}=\frac{1}{\sqrt{N_1}
}\{\alpha(x|000000001\rangle+x|000000010\rangle+y|000000100\rangle)\nonumber\\+\beta(x|000001000\rangle+x|000010000\rangle+y|000100000\rangle)\nonumber\\+\gamma(x|001000000\rangle+x|010000000\rangle+y|100000000\rangle)\}
\end{eqnarray}
where  $N_1$ is the normalization constant,\\
and $x=\frac{\sqrt{2}}{3\sqrt{3}},y=\frac{2\sqrt{2}}{3\sqrt{3}}$.\\
Now if Alice applies $\sigma_z$ operator on her cloned qubits '4'
and '7' and Bob, Charlie apply $\sigma_y$ on their original
qubits 2 and 3 respectively , then the nine qubit entangled state
takes the form,
\begin{eqnarray}
|\bar{X_3}\rangle_{147258369}=\frac{1}{\sqrt{N_1}
}\{\alpha(-x|011100101\rangle-x|011100110\rangle+y|011100000\rangle)\nonumber\\+\beta(-x|011101100\rangle-x|011110100\rangle+y|011000100\rangle)\nonumber\\+\gamma(-x|010100100\rangle-x|001100100\rangle+y|111100100\rangle)\}
\end{eqnarray}
Interestingly we find a five qubit state shared by three parties
given by the density matrix,
\begin{eqnarray}
\rho_{15869}=\frac{1}{N_1}\{x^2\alpha^2
|00001\rangle\langle 00001|+x^2\alpha^2|00001\rangle\langle 00010|+x^2\alpha\beta|00001\rangle\langle 00100|\nonumber\\
+x^2\alpha\beta|00001\rangle\langle
01000|+xy\alpha\gamma|00001\rangle\langle
10000|+x^2\alpha^2|00010\rangle\langle
00010|+x^2\alpha^2|00010\rangle\langle
00001|\nonumber\\+x^2\alpha\beta|00010\rangle\langle 00100
|+x^2\alpha\beta|00010\rangle\langle 01000|
+xy\alpha\gamma|00010\rangle\langle 10000
|\nonumber\\+y^2\alpha^2|00000\rangle\langle 00000
|+x^2\beta^2|00100\rangle\langle 00100
|+x^2\alpha\beta|00100\rangle\langle 00001
|\nonumber\\+x^2\alpha\beta|00100\rangle\langle 00010
|+x^2\beta^2|00100\rangle\langle 01000
|+xy\beta\gamma|00100\rangle\langle 10000
|\nonumber\\+x^2\beta^2|01000\rangle\langle 01000
|+x^2\beta^2|01000\rangle\langle 00100
|+x^2\alpha\beta|01000\rangle\langle
00001|+x^2\alpha\beta|01000\rangle\langle 00010|
\nonumber\\+xy\beta\gamma|01000\rangle\langle 10000
|+y^2\beta^2|00000\rangle\langle 00000
|+x^2\gamma^2|00000\rangle\langle 00000
|\nonumber\\+x^2\gamma^2|00000\rangle\langle 00000
|+y^2\gamma^2|10000\rangle\langle 10000
|+xy\alpha\gamma|10000\rangle\langle 00001
|\nonumber\\+xy\alpha\gamma|10000\rangle\langle 00010
|+xy\beta\gamma|10000\rangle\langle 00100
|+xy\beta\gamma|10000\rangle\langle 01000|\}
\end{eqnarray}
In order to show that the above five qubit states are entangled ,
we have to show that each of the two qubit density matrices
$\rho_{15},\rho_{16},\rho_{58},\rho_{69},\rho_{86}$ , is
entangled. Simultaneously  we must show that the local output
states $\rho_{17},\rho_{14},\rho_{25},\rho_{28},\rho_{36}.\rho_{39}$ are separable.\\
Now, the non local output states are given as,
\begin{eqnarray}
\rho_{15}=\frac{1}{N_1}\{x^2\alpha^2|00\rangle\langle
00|+x^2\alpha^2|00\rangle\langle 00|+y^2\alpha^2|00\rangle\langle
00|\nonumber\\+x^2\beta^2|00\rangle\langle
00|+x^2\beta^2|01\rangle\langle 01|+xy\beta\gamma|01\rangle\langle
10|\nonumber\\+y^2\beta^2|00\rangle\langle
00|+x^2\gamma^2|00\rangle\langle 00|+x^2\gamma^2|00\rangle\langle
00|\nonumber\\+y^2\gamma^2|10\rangle\langle
10|+xy\beta\gamma|10\rangle\langle 01|\}
\end{eqnarray}
\begin{eqnarray}
\rho_{58}=\frac{1}{N_1}\{x^2\alpha^2|00\rangle\langle
00|+x^2\alpha^2|00\rangle\langle 00|+x^2\alpha^2|00\rangle\langle
00|+y^2\alpha^2|00\rangle\langle
00|\nonumber\\+x^2\beta^2|01\rangle\langle
01|+x^2\beta^2|01\rangle\langle 01|+x^2\beta^2|10\rangle\langle 10
|+x^2\beta^2|10\rangle\langle
01|\nonumber\\+y^2\beta^2|00\rangle\langle
00|+x^2\gamma^2|00\rangle\langle 00|+x^2\gamma^2|00\rangle\langle
00|\nonumber\\+y^2\gamma^2|10\rangle\langle
10|+x^2\beta^2|01\rangle\langle 10|\}
\end{eqnarray}
\begin{eqnarray}
\rho_{16}=\frac{1}{N_1}\{x^2\alpha^2|00\rangle\langle
00|+x^2\alpha^2|01\rangle\langle
01|+xy\alpha\gamma|01\rangle\langle
10|\nonumber\\+y^2\alpha^2|00\rangle\langle
00|+x^2\beta^2|00\rangle\langle 00|+x^2\beta^2|00\rangle\langle
00|\nonumber\\+x^2\gamma^2|00\rangle\langle
00|+x^2\gamma^2|00\rangle\langle 00|+y^2\gamma^2|10\rangle\langle
10|+xy\alpha\gamma|10\rangle\langle 01|\}
\end{eqnarray}
\begin{eqnarray}
\rho_{69}=\frac{1}{N_1}\{x^2\alpha^2|01\rangle\langle
01|+x^2\alpha^2|01\rangle\langle 10|+x^2\alpha^2|10\rangle\langle
10|\nonumber\\+x^2\alpha^2|10\rangle\langle
01|+y^2\alpha^2|00\rangle\langle 00|+x^2\beta^2|00\rangle\langle
00|\nonumber\\+x^2\beta^2|00\rangle\langle
00|+x^2\beta^2|00\rangle\langle 00|+y^2\beta^2|00\rangle\langle
00|\nonumber\\+x^2\gamma^2|00\rangle\langle
00|+x^2\gamma^2|00\rangle\langle 00|+y^2\gamma^2|00\rangle\langle
00|\}
\end{eqnarray}
\begin{eqnarray}
\rho_{86}=\frac{1}{N_1}\{x^2\alpha^2|00\rangle\langle
00|+x^2\alpha^2|01\rangle\langle
01|+x^2\alpha\beta|01\rangle\langle
10|\nonumber\\+y^2\alpha^2|00\rangle\langle
00|+x^2\beta^2|10\rangle\langle
10|+x^2\alpha\beta|10\rangle\langle
01|\nonumber\\+x^2\beta^2|00\rangle\langle
00|+y^2\beta^2|00\rangle\langle 00|+x^2\gamma^2|00\rangle\langle
00|+x^2\gamma^2|00\rangle\langle 00|+y^2\gamma^2|00\rangle\langle
00|\}
\end{eqnarray}
Now, $W_4<0$ for
$\rho_{15},\rho_{58},\rho_{16},\rho_{69},\rho_{86}$ for all values
of $\alpha,\beta,\gamma$. Since at least one of $W_3$ and $W_4$ is negative
we can easily infer that the above non local subsystems are entangled. \\
The local output states are given by,
\begin{eqnarray}
\rho_{17}=\rho_{14}=\frac{1}{N_1}\{x^2\alpha^2|01\rangle\langle 01
|+x^2\alpha^2|01\rangle\langle 01 |+y^2\alpha^2|01\rangle\langle
01 |\nonumber\\+x^2\beta^2|01\rangle\langle 01
|+y^2\beta^2|01\rangle\langle 01 |+x^2\gamma^2|01\rangle\langle 01
|\nonumber\\+x^2\gamma^2|00\rangle\langle 00
|+xy\gamma^2|11\rangle\langle 00 |+xy\gamma^2|10\rangle\langle 01
|+y^2\gamma^2|11\rangle\langle 11 |\}
\end{eqnarray}
\begin{eqnarray}
\rho_{28}=\rho_{25}=\frac{1}{N_1}\{-x^2\alpha^2|10\rangle\langle
10 |-x^2\alpha^2|10\rangle\langle 10
|-y^2\alpha^2|10\rangle\langle 10
|\nonumber\\-x^2\beta^2|10\rangle\langle 10
|-x^2\beta^2|11\rangle\langle 11 |+xy\beta^2|11\rangle\langle 00
|+xy\beta^2|00\rangle\langle 11
|\nonumber\\-y^2\beta^2|00\rangle\langle 00
|-x^2\gamma^2|10\rangle\langle 10 |-x^2\gamma^2|10\rangle\langle
10 |-y^2\gamma^2|10\rangle\langle 10 |\}
\end{eqnarray}
\begin{eqnarray}
\rho_{36}=\rho_{39}=\frac{1}{N_1}\{-x^2\alpha^2|10\rangle\langle
10 |-x^2\alpha^2|11\rangle\langle 11 |+xy\alpha^2|11\rangle\langle
00 |\nonumber\\-y^2\alpha^2|00\rangle\langle 00
|+xy\alpha^2|00\rangle\langle 11|-x^2\beta^2|10\rangle\langle
10|\nonumber\\-x^2\beta^2|10\rangle\langle
10|-y^2\beta^2|10\rangle\langle 10|-x^2\gamma^2|10\rangle\langle
10|-x^2\gamma^2|10\rangle\langle 10|y^2\beta^2|10\rangle\langle
10|
\end{eqnarray}
Now $W_4=W_3=0$ for $\rho_{14},\rho_{17}$ and $W_4>0,W_3>0$ for
$\rho_{25},\rho_{28},\rho_{36},\rho_{39}$ clearly indicating the
fact that all the local output states are separable.\\
Though in this work we have considered particular cases and not
dealt with all possible measurement results, we have achieved our
target of secretly generating five qubit entangled state. One can
investigate other measurement results to see how they can be used
in other broadcasting problems. This five qubit entangled state
will play crucial role as a secret quantum channel between three
distant partners.
\section{Conclusion:}
To summarize, the motivation of our work which was to broadcast
five qubit entangled state among three parties secretly. This can
be done by applying twice B-H quantum cloning machine followed by
subsequent measurement on machine state vectors by three parties.
Not only that each of these parties apply local operations after
the cloning transformation to make the local output states
separable. Our problem is interesting in the sense that this is
the first attempt to broadcast five qubit entangled state
secretly. This state can be used by Alice,Bob, Charlie as a five
qubit entangled state in various cryptographic protocols viz.
quantum key distribution protocols. It is also interesting in the
sense that in this protocol we have broadcasted five qubit
entangled state in such a way that it is independent of input
parameters $\alpha,\beta,\gamma$.
\section{Acknowledgement}
I.C acknowledges almighty God for being the source of inspiration
of all work. He also acknowledges Prof C.G.Chakraborti for being
the source of inspiration in research work.
\section{References}
$[1]$ W.K.Wootters, W.H.Zurek, Nature 299 (1982) 802.\\ $[2]$
V.Buzek, M.Hillery, Phys.Rev.A 54 (1996) 1844.\\ $[3]$ N.Gisin,
S.Massar, Phys.Rev.Lett. 79 (1997) 2153.\\ $[4]$ V.Buzek,
S.L.Braunstein, M.Hillery, D.Bruss, Phys.Rev.A 56 (1997) 3446.\\
$[5]$ D.Bruss, D.P.DiVincenzo, A.Ekert, C.A.Fuchs, C.Macchiavello,
J.A.Smolin,Phys.Rev.A 57 (1998) 2368.\\
$[6]$ A.K.Pati, S.L.Braunstein, Nature 404 (2000) 164.\\
$[7]$ L.M.Duan and G.C.Guo,Phys.Rev.Lett. 80 (1998)4999.\\
$[8]$ Einstein, Podolsky and Rosen, Phys.Rev. 47 (1935)777.\\
$[9]$ C.H.Bennett and G.Brassard, Proceedings of IEEE
International Conference on Computers, System and Signal
Processing, Bangalore, India, 1984, pp.175-179.\\
$[10]$ P.W.Shor and J.Preskill, Phys.Rev.Lett. 85 (2000)441.\\
$[11]$ C.H.Bennett and S.J.Weisner, Phys.Rev.Lett.69 (1992)2881.\\
$[12]$ C.H.Bennett, G.Brassard, C.Crepeau, R.Jozsa, A.Peres
and W.K.Wootters,Phys.Rev.Lett. 70 (1993)1895.\\
$[13]$ S.Bandyopadhyay, G.Kar, Phys.Rev.A 60 (1999)3296.\\
$[14]$ I.Ghiu, Phys.Rev.A 67 (2003)012323.\\
$[15]$ S.Adhikari, B.S.Choudhury and I.Chakrabarty, J. Phys. A: Math. Gen. 39 No 26 (2006)8439.\\
$[16]$ S.Adhikari, B.S.Choudhury, Phys. Rev. A74,  (2006)032323.\\
$[17]$   Satyabrata Adhikari, A. S. Majumdar, N. Nayak
, arXiv:0708.1869.\\
$[18]$  Indranil Chakrabarty, B.S.Choudhury, arXiv:0803.3393.\\
$[19]$ P.W.Shor and J.Preskill, Phys.Rev.Lett. 85 (2000)441.\\
$[20]$ L.Goldenberg and L.Vaidman, Phys.Rev.Lett. 75 (1995)1239.\\
$[21]$ A.Peres, Phys.Rev.Lett. 77 (1996)1413.\\
$[22]$ M.Horodecki, P.Horodecki, R.Horodecki, Phys.Lett.A 223.
(1996)1.\\
\end{document}